\documentclass[a4paper,conference]{IEEEtran}
\DeclareRobustCommand*{\IEEEauthorrefmark}[1]{%
  \raisebox{0pt}[0pt][0pt]{\textsuperscript{\footnotesize\ensuremath{#1}}}}
\IEEEoverridecommandlockouts
\usepackage{amsfonts} 
\usepackage{tikz} 
\usetikzlibrary{bayesnet,matrix} 
\usepackage{enumerate} 
\usepackage{array} 
\usepackage{pgfplots} 
\usepackage{multirow}
\ifCLASSINFOpdf
\else
\fi
\begin{document}
%
\title{Probabilistic Latent Factor Model for Collaborative Filtering with Bayesian Inference}


%
\author{\IEEEauthorblockN{Jiansheng Fang\IEEEauthorrefmark{1,2,3}, Xiaoqing Zhang\IEEEauthorrefmark{2}, Yan Hu\IEEEauthorrefmark{2}, Yanwu Xu\IEEEauthorrefmark{4}, Ming Yang\IEEEauthorrefmark{3}, Jiang Liu\IEEEauthorrefmark{*2,4}}
\thanks{$^{*}$ Corresponding author: Jiang Liu}
\IEEEauthorblockA{\IEEEauthorrefmark{1}School of computer science and technology, Harbin Institute of Technology, Harbin 150001, China\\ Email: 11949039@mail.sustech.edu.cn}
\IEEEauthorblockA{\IEEEauthorrefmark{2}Department of computer science and engineering, Southern University of Science and Technology, Shenzhen
518055, China\\Email: liuj@sustech.edu.cn}
\IEEEauthorblockA{\IEEEauthorrefmark{3}CVTE Research, Guangzhou 510530, China\\Email: yangming@cvte.com}
\IEEEauthorblockA{\IEEEauthorrefmark{4}Cixi Institute of Biomedical Engineering, Chinese Academy of Sciences, Ningbo 315201, China}}

\maketitle

\begin{abstract}
Latent Factor Model (LFM) is one of the most successful methods for Collaborative filtering (CF) in the recommendation system, in which both users and items are projected into a joint latent factor space. Base on matrix factorization applied usually in pattern recognition, LFM models user-item interactions as inner products of factor vectors of user and item in that space and can be efficiently solved by least square methods with optimal estimation. However, such optimal estimation methods are prone to overfitting due to the extreme sparsity of user-item interactions. In this paper, we propose a Bayesian treatment for LFM, named Bayesian Latent Factor Model (BLFM). Based on observed user-item interactions, we build a probabilistic factor model in which the regularization is introduced via placing prior constraint on latent factors, and the likelihood function is established over observations and parameters. Then we draw samples of latent factors from the posterior distribution with Variational Inference (VI) to predict expected value. We further make an extension to BLFM, called BLFMBias, incorporating user-dependent and item-dependent biases into the model for enhancing performance. Extensive experiments on the movie rating dataset show the effectiveness of our proposed models by compared with several strong baselines.
\end{abstract}


%
\IEEEpeerreviewmaketitle

\section{Introduction}
One of the popular approaches to recommender systems \cite{gomez2015netflix,ricci2011introduction} which meet users' personalized needs for items is Collaborative filtering (CF) \cite{wang2019sequential,glauber2019collaborative,yang2014survey,shi2014collaborative,linden2003amazon,sarwar2001item} which heavily relies on the past user behavior (e.g., ratings and clicks). A major appeal of CF is that CF techniques require no domain knowledge and avoid the need for extensive data collection \cite{koren2008factorization,koren2009matrix}. Many online services have widely adopted personalized recommendation to enhance user satisfaction. One of the most successful realizations to CF is Latent Factor Model (LFM) in which both user and item are projected into a joint latent factor space. The main idea behind LFM is to model user-item interactions as inner products of factor vectors of user and item in the space.

LFM is capable of learning a joint latent factor space from user-item interactions which include high-quality explicit rating \cite{salakhutdinov2007restricted} and abundant implicit feedback \cite{hu2008collaborative,he2016fast,lake2019large}. Due to scalability and flexibility of LFM \cite{billsus1998learning}, much research effort have been devoted to enhancing its recommendation effect \cite{mongia2020deep,sun2020multi,huang2019gaussian}. For example, LFM incorporates side information into the user-item interaction function to improve performance \cite{verma2017collaborative,sun2019research,he2016vbpr}. Furthermore, probabilistic factor models have been proposed to overcome the sparseness and solve the non-convex optimization problem \cite{chen2019data,xiong2010temporal,mnih2008probabilistic,marlin2004modeling}. 

Conventional probabilistic factor models \cite{da2019prior,zhao2015bayesian,lim2007variational,salakhutdinov2008bayesian} construct the likelihood function over the rating matrix. However, in our model, the likelihood function is given over only observed user-item interactions, and the benefit is that we can incorporate additional information into the model, such as biases, attributes of a user. Both of additional information and latent factors are placed prior constraint for each user and item. Because the posterior distribution is hard to exact inference, several probabilistic factor models use Markov chain Monte Carlo (MCMC) \cite{yang2019parallelized,gong2019icebreaker,gautier2017zonotope} to conduct approximate posterior inference. MCMC provides guarantee of producing exact samples from the target distribution. An alternative method of approximate posterior inference is Variational Inference (VI) which tends to be faster and easier to scale to large data \cite{meng2019variational,blei2017variational,jaakkola2000bayesian}. Considering run efficiency, we adopt VI to draw samples from the posterior distribution.

Training LFM amounts to fit the inner products of factor vectors to observed user-item interactions over the given loss function. Optimal estimation methods are widely used to estimate model parameters, such as Maximum Likelihood (ML) \cite{stigler2007epic}, Maximum-A-Posteriori (MAP) \cite{bassett2019maximum}. Although optimal estimation methods can be worked efficiently on large datasets, they are easy to cause overfitting due to extreme sparsity of user-item interactions. Hence we explore the Bayesian treatment of LFM based on observed user-item interactions to alleviate overfitting. The main contributions of this work are as follows:
\begin{itemize}
\item We propose a Bayesian treatment for LFM, named Bayesian Latent Factor Model (BLFM), in which a probabilistic factor model is given over latent factors and observed data, and the values of model parameters are drawn from posterior distribution.
\item We further make an extension to BLFM, called BLFMBias, incorporating user-dependent and item-dependent biases into model to enhance performance. This shows that BLFM can be extended to incorporate additional information to solve cold start problem in the recommendation system. 
\item Extensive experiments are performed on the movie rating dataset to demonstrate the effectiveness of our proposed method.
\end{itemize}

\section{Problem Statement}
In this section, we formalize the problem and introduce optimal estimation method for LFM.

\subsection{Rating Prediction}
Let \begin{math}m\end{math} and \begin{math}n\end{math} denote the number of users and items, respectively. Given observed rating \begin{math}r_{ui}\end{math} of user \begin{math}u\end{math} on item \begin{math}i\end{math}, let \begin{math}\textbf{D}\end{math} denotes the set of tuple (\begin{math}u\end{math}, \begin{math}i\end{math}, \begin{math}r_{ui}\end{math}). The recommendation problem is formulated as a problem of predicting unobserved rating. Formally, the problem of rating prediction is to learn 
\begin{equation}\hat{r}_{ui}= f(r_{ui}|\Theta)\label{eq:1},\end{equation} 
where \begin{math}\hat{r}_{ui}\end{math} denotes the estimated rating of user \begin{math}u\end{math} on item \begin{math}i\end{math}, and \begin{math}\Theta\end{math} denotes model parameters. The function \begin{math}f\end{math} establishes the link between model parameters and observed rating.

\subsection{Optimal Estimation}
In LFM, rating is modeled as inner product of factor vectors with dimension \begin{math}K\end{math}. Let \begin{math}\textbf{p}_{u}\end{math} and \begin{math}\textbf{q}_{i}\end{math} denote the factor vector for user \begin{math}u\end{math} and item \begin{math}i\end{math}, respectively. Equation (\ref{eq:1}) can be recast as follow: 
\begin{equation}
\hat{r}_{ui}=f(r_{ui}|\textbf{p}_{u}, \textbf{q}_{i})=\textbf{p}_{u}^{T}\textbf{q}_{i}=\sum_{k=1}^{K}p_{uk}q_{ik}\label{eq:2},
\end{equation}
where \begin{math}k\end{math} is the dimension of factor vector. Training LFM is to fit the inner products of factor vectors \begin{math}\textbf{p}_{u}\end{math} and \begin{math}\textbf{q}_{i}\end{math} to the observed rating \begin{math}r_{ui}\end{math}.

The optimal estimation method learns the model parameters of LFM by minimizing the regularized squared error 
\begin{equation}
\min \limits_{\textbf{p}_{*},\textbf{q}_{*}} \sum_{(u,i,r_{ui})\in \textbf{D}} (r_{ui}- \textbf{p}_{u}^{T}\textbf{q}_{i})^{2} + \lambda (\|\textbf{p}_{u}\|^{2} + \|\textbf{q}_{i}\|^{2})\label{eq:3},
\end{equation}
where \begin{math}\lambda \end{math} is the regularization parameter. Given the loss function of Equation (\ref{eq:3}), gradient descent is usually applied to optimize parameters.

Such optimal estimation method can be interpreted as MAP. MAP can be seen as a regularization of ML estimation, both of MAP and ML can be used to make single optimal estimate over observed data. However, the regularization parameters should be tuned carefully to overcome overfitting. For improving generalization performance, we apply Bayesian parameter estimation method to estimate factor vectors \begin{math}\textbf{p}_{u}\end{math} and \begin{math}\textbf{q}_{i}\end{math}.

\section{Bayesian Latent Factor Model (BLFM)}
In this section, we present the details of BLFM and its extend BLFMBias. First, we define a probabilistic factor model, in which prior assumption is introduced for model parameters and likelihood function is build on observations and parameters. Second, based on the defined probabilistic factor model, we draw samples from posterior distribution of model parameters to estimate the expected rating. 

\subsection{Prior Assumption}
In fact, we recast Equation (\ref{eq:3}) by building a probabilistic factor model with observed rating being observations, factor vectors and biases terms being parameters. In this model, we introduce regularization by placing prior constraint over model parameters.

\textbf{BLFM.} 
The prior distributions over the factor vectors are assumed to be Gaussian: 
\begin{equation}
\begin{array}{ll}
p(p_{uk}| \mu_{uk}, \sigma_{uk}) \sim \mathcal{N}(\mu_{uk}, \sigma_{uk}),\\
\\
p(q_{ik}| \mu_{ik}, \sigma_{ik}) \sim \mathcal{N}(\mu_{ik}, \sigma_{ik}).
\end{array} \label{eq:4}
\end{equation}
Here, the \begin{math}k^{th}\end{math} dimension of factor vector have its independent hyperparameters, mean \begin{math}\mu_{uk}\end{math}, \begin{math}\mu_{ik}\end{math} and variance \begin{math}\sigma_{uk}\end{math}, \begin{math}\sigma_{ik}\end{math}. Such that amounts to set the weight of each latent factor. It is flexible enough to control the complexity of model and adjust the effect of regularization. 

\textbf{BLFMBias.} 
The biases tend to capture the vibration of explicit rating of user on item. We incorporate the biases terms of user and item into model to denoise. Let \begin{math}\overline{r}\end{math} denotes the overall average rating, the prior for bias \begin{math}{b}_{u}\end{math} of user \begin{math}u\end{math}, bias \begin{math}{b}_{i}\end{math} of item \begin{math}i\end{math} in BLFMBias are defined as:
\begin{equation}
\begin{array}{ll}
p(b_{u}|\overline{r}) \sim \mathcal{N}(0, \overline{r}),\\
\\
p(b_{i}|\overline{r}) \sim \mathcal{N}(0, \overline{r}).\\
\end{array}\label{eq:5}
\end{equation}
The parameters \begin{math}{b}_{u}\end{math} and \begin{math}{b}_{i}\end{math} indicate the observed deviations of user \begin{math}u\end{math} and item \begin{math}i\end{math}. Thus, the biases terms is defined as:
\begin{equation}
b_{ui} = \overline{r} + b_{u} + b_{i}\label{eq:6}.
\end{equation}

\subsection{Likelihood Function}
The likelihood function establishes the relation between observations and parameters. 

\begin{figure}[htbp]
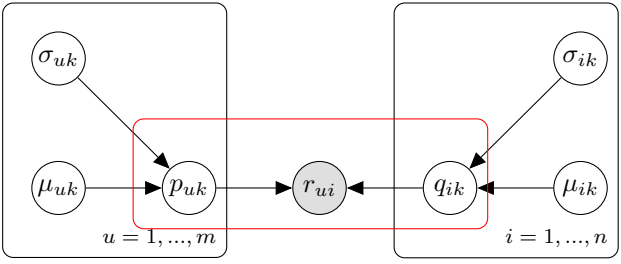

  \centering
  \tikz{
    \node[latent] (umu) {$\mu_{uk}$} ; 
    \node[latent, above=of umu] (usigma) {$\sigma_{uk}$} ; 
    \node[latent, right=of umu] (puk) {$p_{uk}$} ; 
    \node[obs, right=of puk] (rui) {$r_{ui}$} ; 
    \node[latent, right=of rui] (qik) {$q_{ik}$} ;
    \node[latent, right=of qik] (imu) {$\mu_{ik}$} ;
    \node[latent, above=of imu] (isigma) {$\sigma_{ik}$} ;
    \edge {umu, usigma} {puk} ; 
    \edge {imu, isigma} {qik} ;
    \edge {puk,qik} {rui} ;
    \plate[inner sep=0.25cm, xshift=-0.12cm, yshift=0.12cm] {plate1} {(umu) (usigma) (puk)} {$u=1,...,m$};
    \plate[inner sep=0.25cm, xshift=-0.12cm, yshift=0.12cm] {plate2} {(imu) (isigma) (qik)} {$i=1,...,n$};
    \plate[inner sep=0.25cm, xshift=-0.12cm, yshift=0.30cm, color=red] {plate3} {(puk) (qik) (rui)} {};
  }
  \caption{The graphical model of BLFM}
  \label{fig:1}
\end{figure}

\textbf{BLFM.}
In BLFM, we fit the inner product of factor vectors to approximate the observed rating. The likelihood function of BLFM is defined as:  
\begin{equation}
p(r_{ui}|\textbf{p}_{u}, \textbf{q}_{i}) \sim \mathcal{N}(\sum_{k=1}^{K}p_{uk}q_{ik}, \overline{r}) \label{eq:7}.
\end{equation}
A graphical model corresponding to this generative process is shown at the Figure \ref{fig:1}.

\begin{figure}[htbp]
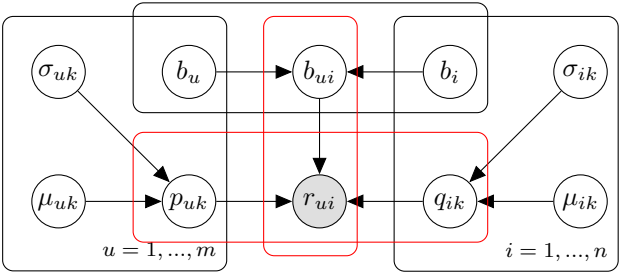

  \centering
  \tikz{
    \node[latent] (umu) {$\mu_{uk}$} ; 
    \node[latent, above=of umu] (usigma) {$\sigma_{uk}$} ; 
    \node[latent, right=of umu] (puk) {$p_{uk}$} ; 
    \node[obs, right=of puk] (rui) {$r_{ui}$} ; 
    \node[latent, right=of rui] (qik) {$q_{ik}$} ;
    \node[latent, right=of qik] (imu) {$\mu_{ik}$} ;
    \node[latent, above=of imu] (isigma) {$\sigma_{ik}$} ;
    \node[latent, above=of rui] (bui) {$b_{ui}$} ;
    \node[latent, left=of bui] (bu) {$b_{u}$} ;
    \node[latent, right=of bui] (bi) {$b_{i}$} ;
    \edge {umu, usigma} {puk} ; 
    \edge {imu, isigma} {qik} ;
    \edge {puk,qik, bui} {rui} ;
    \edge {bu,bi} {bui} ;
    \plate[inner sep=0.25cm, xshift=-0.12cm, yshift=0.12cm] {plate1} {(umu) (usigma) (puk)} {$u=1,...,m$};
    \plate[inner sep=0.25cm, xshift=-0.12cm, yshift=0.12cm] {plate2} {(imu) (isigma) (qik)} {$i=1,...,n$};
    \plate[inner sep=0.25cm, xshift=-0.12cm, yshift=0.30cm, color=red] {plate3} {(puk) (qik) (rui)} {};
    \plate[inner sep=0.25cm, xshift=-0.12cm, yshift=0.30cm] {plate4} {(bu) (bi) (bui)} {};
    \plate[inner sep=0.25cm, xshift=-0.12cm, yshift=0.12cm, color=red] {plate5} {(bui) (rui)} {};
  }
  \caption{The graphical model of BLFMBias.}
  \label{fig:2}
\end{figure}

\textbf{BLFMBias.}
In BMFBias, we fit the sum of inner product of factor vectors and biases terms to the observed rating. The likelihood function of BMFBias is defined as: 
\begin{equation}
p(r_{ui}|\textbf{p}_{u}, \textbf{q}_{i}, b_{u}, b_{i}) \sim \mathcal{N}(\sum_{k=1}^{K}p_{uk}q_{ik} + b_{ui} , \overline{r}).
\end{equation}
A graphical model corresponding to this generative process is shown at the Figure \ref{fig:2}.

\subsection{Variational Inference (VI)}
While exact inference of the posterior distribution over the parameters given the observed rating is intractable, we resort to use VI for posterior approximate inference. The goal of VI is to approximate a conditional density of latent variables given observed variables \cite{beal2003variational}. VI tries to find a family of approximate density and turns the inference problem into an optimization problem by minimizing the Kullback-Leibler (KL) divergence to the exact posterior.

\textbf{BLFM.}
The KL divergence is an information-theoretical measure of proximity between two density \cite{perez2008kullback}. First, Let us give the posterior density \begin{math}p(\textbf{p}_{u},\textbf{q}_{i}|\textbf{D})\end{math} of BLFM. Then, we specify a family \begin{math}Q\end{math} over the factor vectors \begin{math}\textbf{p}_{u}\end{math}, \begin{math}\textbf{q}_{i}\end{math}, and each \begin{math}q(z) \in Q\end{math} is a candidate approximation to the exact posterior density. Last, our purpose is to solve the following optimization:
\begin{equation}
q^{*} = \mathop{\arg\min}_{q(\textbf{p}_{u},\textbf{q}_{i}) \in Q} KL(q(\textbf{p}_{u},\textbf{q}_{i}) \| p(\textbf{p}_{u},\textbf{q}_{i}|\textbf{D})).
\end{equation}
Within the family \begin{math}Q\end{math}, the goal is to find the best approximation  \begin{math}q^{*}(\textbf{p}_{u},\textbf{q}_{i})\end{math} of the posterior \begin{math}p(\textbf{p}_{u},\textbf{q}_{i}|\textbf{D})\end{math}. KL divergence is the expected value of the difference between target density and candidate density, the expectation representation of KL can be wirten as:
\begin{equation}
\begin{array}{ll}
KL(q(\textbf{p}_{u},\textbf{q}_{i}) \| p(\textbf{p}_{u},\textbf{q}_{i}|\textbf{D})) = \\
\\
\mathbb{E}[\log q(\textbf{p}_{u},\textbf{q}_{i})] - \mathbb{E}[\log p(\textbf{p}_{u},\textbf{q}_{i}|\textbf{D})].
\end{array}
\end{equation}

According to the Bayesian formula, the posterior (conditional) density  \begin{math}p(\textbf{p}_{u},\textbf{q}_{i}|\textbf{D})\end{math} can be represented as:
\begin{equation}
p(\textbf{p}_{u},\textbf{q}_{i}|\textbf{D}) = \frac{p(\textbf{p}_{u},\textbf{q}_{i},\textbf{D})}{p(\textbf{D})}.
\end{equation}
Here, the numerator \begin{math}p(\textbf{p}_{u},\textbf{q}_{i},\textbf{D})\end{math} is a joint density based on the factor vectors and the observed rating, and the denominator \begin{math}p(\textbf{D})\end{math} contains the marginal density of the observations, also called the evidence. This evidence is calculated by marginalizing out the latent variables from the joint density, and this evidence integral is intractable. 

Because the evidence is hard to compute, VI optimize an alternative objective function: 
\begin{equation}
ELBO(q) = \mathbb{E}[\log p(\textbf{p}_{u},\textbf{q}_{i},\textbf{D})] - \mathbb{E}[\log q(\textbf{p}_{u},\textbf{q}_{i})]\label{eq:12}.
\end{equation}
This function is called the evidence lower bound (ELBO). Maximizing the ELBO is equivalent to minimizing the KL divergence.

According to the Bayesian formula, the joint density \begin{math}p(\textbf{p}_{u},\textbf{q}_{i},r_{ui})\end{math} can be represented as:
\begin{equation}
p(\textbf{p}_{u},\textbf{q}_{i}, \textbf{D}) = p(\textbf{D}|\textbf{p}_{u},\textbf{q}_{i})p(\textbf{p}_{u})p(\textbf{q}_{i}).
\end{equation}
Here, \begin{math}p(\textbf{D}|\textbf{p}_{u},\textbf{q}_{i})\end{math} denotes likelihood function, and the \begin{math}p(\textbf{p}_{u})\end{math}, \begin{math}p(\textbf{q}_{i})\end{math} denote the prior distribution of \begin{math}\textbf{p}_{u}\end{math}, \begin{math}\textbf{q}_{i}\end{math}, respectively.

With integrating the prior distribution (\ref{eq:4}), the likelihood function (conditional density) (\ref{eq:7}), and the ELBO function (\ref{eq:12}), we can approximate posterior over the factor vectors \begin{math}\textbf{p}_{u}\end{math}, \begin{math}\textbf{q}_{i}\end{math} given the observed rating \begin{math}\textbf{D}\end{math}. Then we draw samples of model parameters from the posterior \begin{math}p(\textbf{p}_{u},\textbf{q}_{i}|\textbf{D})\end{math}. Given a pair \begin{math}(u,i)\end{math}, the expected value of rating \begin{math}\hat{y}_{ui}\end{math} is predicted as:
\begin{equation}
\hat{r}_{ui} = \int_{\textbf{p}_{u},\textbf{q}_{i}} p({r}_{ui}|\textbf{p}_{u},\textbf{q}_{i})p(\textbf{p}_{u},\textbf{q}_{i}|\textbf{D}) \mathrm{d} \{\textbf{p}_{u},\textbf{q}_{i}\}.
\end{equation}

\textbf{BLFMBias.}
In BLFMBias, we incorporate the parameters \begin{math}{b}_{u}\end{math} and \begin{math}{b}_{i}\end{math} into Bayesian model. We use \begin{math}\Theta\end{math} unified denotes the parameters of posterior density in BLFMBias, including \begin{math}\textbf{p}_{u}\end{math}, \begin{math}\textbf{q}_{i}\end{math},  \begin{math}b_{u}\end{math}, \begin{math}b_{i}\end{math}. The objective function of VI is:
\begin{equation}
ELBO(q) = \mathbb{E}[\log p(\Theta,\textbf{D})] - \mathbb{E}[\log q(\Theta)].
\end{equation}
Here, our aim is to find the best \begin{math}q(\Theta)\end{math} to approximate the posterior density \begin{math}p(\Theta|\textbf{D})\end{math} by maximizing the ELBO. And the joint density \begin{math}p(\Theta,\textbf{D})\end{math} of BLFMBias can be represented as:
\begin{equation}
p(\Theta,\textbf{D}) = p(\textbf{D}|\Theta)p(\Theta).
\end{equation}
Here, \begin{math}p(\Theta)\end{math} is the prior distribution of parameters in BMFBias, and \begin{math}p(\textbf{D}|\Theta)\end{math} is the likelihood function given the observed rating. As same with BLFM, we can predict expectation rating for user \begin{math}u\end{math} and item \begin{math}i\end{math}:
\begin{equation}
\hat{r}_{ui} = \int_{\Theta} p({r}_{ui}|\Theta)p(\Theta|\textbf{D}) \mathrm{d} \Theta.
\end{equation}

The posterior probability is a conditional probability conditioned on observed rating. For BLFM, the parameter of posterior probability is latent factors. For BLFMBias, the parameter of posterior probability is latent factors and biases terms. By learning probabilistic factor model, we draw samples of parameters from posterior probability to predict expected value of rating.

\section{Experiments}
In this section, the following research questions will be answered by conducting corresponding experiments.
\begin{description}
\item[\textbf{RQ1}] Do our proposed BLFM outperforms the state-of-the-art methods for rating prediction? 
\item[\textbf{RQ2}] By incorporating biases into BLFM, how is our proposed BLFMBias performance?
\item[\textbf{RQ3}] Is Bayesian treatment more robust in alleviating overfitting?
\item[\textbf{RQ4}] Do the samples of parameter from posterior probability converge by applying VI?
\item[\textbf{RQ5}] Which hyperparameter most affect performance for training BLFM model?
\end{description}
In what follows, we first present the experimental settings, followed by answering the above five research question.

\subsection{Experimental Settings}
\begin{table}[htbp]
\caption{Statistics of the evaluation datasets.}
\begin{center}
\begin{tabular}{ |c|c|c|c|c| } 
\hline 
\textbf{Dataset} & \textbf{Interaction} & \textbf{User} & \textbf{Item} & \textbf{Sparsity} \\
\hline 
MovieLens & 1,000,209 & 6,040 & 3,706 & 95.53\% \\
\hline 
\end{tabular}
\end{center}\label{Tb:1}
\end{table}
\textbf{Dataset.} We experimented with one public accessible dataset:  MovieLens\footnote[1]{https://grouplens.org/datasets/movielens/1m/}. The characteristic of the dataset are summarized in Table \ref{Tb:1}. This movie rating dataset has been widely used to evaluate LFM algorithms. We choose the version containing about one million ratings, where each user has at least 20 ratings. The ratings are given on a scale of 0 to 5 stars with increments of 0.5 stars. The mean rating over all users and movies is 3.58 and the variance is 1.24.

\textbf{Evaluation Metric.}
We adopted Root Mean Square Error (RMSE) to evaluate the performance of rating prediction. RMSE is widely used in literature \cite{harvey2011bayesian,lim2007variational,salakhutdinov2008bayesian,mnih2008probabilistic}. For each user, we held out his latest interaction as the test set and utilized the remaining data for training. 

\textbf{Baselines.}
In order to evaluate the utility of our models, four strong LFM baselines were selected as follows:
\begin{itemize}
\item[-] \textbf{SVD}\cite{koren2008factorization}. SVD trains model directly only on the observed ratings while avoiding overfitting through introducing regularization, and applied gradient descent to solve the optimization problem. 
\item[-] \textbf{PMF}\cite{mnih2008probabilistic}. Probabilistic Matrix Factorization (PMF) adopts a probabilistic linear model with Gaussian observation noise, and performed well on very sparse and imbalanced datasets by modeling the user preference matrix as as product of two lower-rank matrices. 
\item[-] \textbf{BPMF}\cite{salakhutdinov2008bayesian}. Bayesian Probabilistic Matrix Factorization (BPMF) implements PMF by using MCMC and further placed Gaussian-Wishart priors on the users and items feature vectors. 
\item[-] \textbf{SVDBias}\cite{koren2008factorization}. SVDBias is a fully-regularized gradient descent SVD model with user and item biases providing a competitive baseline.  
\end{itemize}

Given observed user-item interactions, both of our BLFM and BLFMBias make rating prediction by drawing samples of parameters from posterior probability. 

\textbf{Parameters Setting.}
For SVD and SVDBias, weights are updated using a learning rate of 0.001, the regularization parameter is set to 0.01, and the prediction errors on train set were observed to stabilize after approximately 30 iterations. For PMF model, we set learning rate of 0.005, a momentum of 0.9, and max iterations of 200. For BPMF model, we initialized the two lower-rank matrices of zero, the mean of Gaussian-Wishart priors of zero, the degrees of freedom (parameter of Wishart distribution) of latent factor number, and max iterations of 100. We implemented our proposed methods based on Python\footnote[2]{https://github.com/fjssharpsword/fjscode2019/tree/master/BMF}. We fixed the parameters of Gaussian priors, the mean of zero and the variance of the average rating. On conducting VI, we fit target density 10000 times, and 2000 samples are drawn to calculate expectation rating. For all methods, we tested in different number of latent factors (8, 16, 32, 64) on the same experimental device and reported the best performance given corresponding parameters. 

\subsection{Performance Comparison (RQ1)}
Table \ref{Tb:2} shows the performance of RMSE with respect to the number of latent factors. On the whole, we can see that BLFM achieves the best performance, significantly outperforming SVD and PMF, slightly outperforming BPMF. 
\begin{table}[htbp]
\renewcommand\arraystretch{1.5}
\caption{Performance of RMSE.}
\begin{center}
\begin{tabular}{ |c|c|c|c|c| } 
\hline \multicolumn{5}{|c|}{\textbf{Movielens}} \\
\hline 
\textbf{Factors} & \textbf{SVD} & \textbf{PMF} & \textbf{BPMF} & \textbf{\textit{BLFM}} \\
\hline 
RMSE@8 & 1.0274 & 1.0256 & 0.9871 & \underline{0.9781} \\
RMSE@16 & 1.0256 & 1.0076 & 0.9840 & \underline{0.9836} \\
RMSE@32 & 1.0278 & 0.9999 & 0.9875 & \underline{0.9852} \\
RMSE@64 & 1.0141 & 0.9952 & 0.9867 & \underline{0.9826} \\
\hline 
\end{tabular}
\end{center}\label{Tb:2}
\end{table}

\textbf{SVD.} SVD and BLFM are trained directly on observed rating. Let us observe the performance on factors of 32. BMF model achieves an RMSE of 0.9852, compared to an RMSE of 1.0278 of SVD model, an improvement of over 4.1\%. The result demonstrates that Bayesian prior constraint is better than parameter regularization and the expectation rating prediction is better than the single optimal rating prediction.

\textbf{PMF.} PMF model is trained by maximizing the log-posterior over the users and items features with fixed hyperparameters (i.e. the observation noise variance and prior variance) \cite{salakhutdinov2008bayesian}. PMF model introduced prior constraint and learned model parameters over the rating matrix (included unobserved entries of rating matrix) with MAP method. Let us observe the performance on factors of 16. BLFM achieves an RMSE of 0.9836, compared to an RMSE of 1.0076 of PMF model, an improvement of over 2.38\%.

\textbf{BPMF.} BPMF model adopted fully Bayesian treatment of the PMF model and used MCMC to train model over the rating matrix. Compared to BPMF, BLFM applies Bayesian treatment in the same, but uses the VI to train model over the observed rating. Although BLFM only slightly outperforms BPMF, but it is faster than BPMF and is flexible to extend model. On training latent factors of 64, the run time of BPMF is over 10,000 seconds with executing sampler of 100 step, and BLFM approximates 4,000 seconds with fitting target posterior density of 10000 times. On average, the run time of training BPMF is double longer than BLFM. 

\subsection{Performance with Adding Biases (RQ2)} 
Table \ref{Tb:3} shows the performance of RMSE with adding biases on different latent factors. Let us observe the performance on factors of 16. BLFMBias achieves an RMSE of 0.9397, compared to an RMSE of 0.9836 of BLFM, an improvement of over 4.46\%. This result demonstrate that the performance can be improved by adding biases. Compared to an RMSE of 0.9570 of SVDBias model, BLFMBias also have an improvement of over 1.80\%. 
\begin{table}[htbp]
\renewcommand\arraystretch{1.5}
\caption{Performance of RMSE with adding biases.}
\begin{center}
    \begin{tabular}{ |c|c|c|c| } 
    \hline \multicolumn{4}{|c|}{\textbf{Movielens}} \\
    \hline 
    \textbf{Factors} & \textbf{BLFM} & \textbf{SVDBias} & \textbf{\textit{BLFMBias}} \\
        \hline 
        RMSE@8 & 0.9781 & 0.9581 & \underline{0.9406}  \\
        RMSE@16 & 0.9836 & 0.9570 & \underline{0.9397}  \\
        RMSE@32 & 0.9852 & 0.9573 & \underline{0.9410}  \\
        RMSE@64 & 0.9826 & 0.9572 & \underline{0.9452}  \\
        \hline 
    \end{tabular}
\end{center}\label{Tb:3}
\end{table}

\subsection{Performance of Alleviating Overfitting (RQ3)} 
\begin{figure}[htbp]
  \centering
  \includegraphics[width=\linewidth]{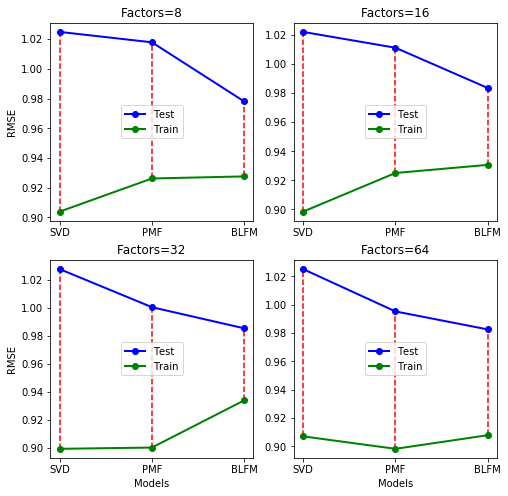}
\caption{Performance of RMSE on validation set and test set.}
  \label{fig:3}
\end{figure}  
To make a comparison between the performance of the train set and testing set, we randomly select rating entries of the same size as the test set from the train set. Figure \ref{fig:3} shows that SVD and PMF achieves better on the validation set (green line) but worse on the test set (blue line), compared to BLFM. The length of dashed red line in the Figure \ref{fig:3} denotes the difference size of performance between validation set and test set. It can be concluded that the length of BLFM is shorter than PMF and the length of SVD is longer than PMF. On alleviating overfitting, this experiment clearly demonstrates that BLFM which applies prior constraint and Bayesian estimation method is best, next is PMF which applies prior constraint and MAP estimation method, and last is SVD which applies parameter regularization and optimal estimation method. This shows that BLFM with prior constraint and Bayesian estimation method is more robust against overfitting.

\subsection{Convergence of Parameter Estimation (RQ4)} 
\begin{figure}[htbp]
  \centering
  \includegraphics[width=\linewidth]{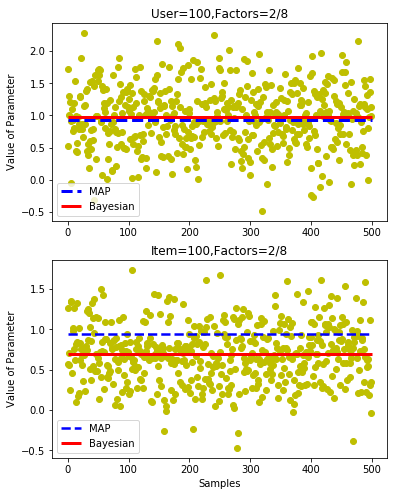}
\caption{Estimation of Bayesian and MAP.}
  \label{fig:4}
\end{figure}

In BLFM, the value of parameters (latent factors) is drawn from approximating posterior distribution. As shown in Figure \ref{fig:4}, we draw 500 samples (x-axis) from the approximation distribution for user 100 and item 100, respectively. We observe the value (y-axis) distribution (yellow circle point) of second dimension of latent factors 8, and calculate the average of these samples. We can see that the value between this mean (solid red line) and MAP optimized point (dashed blue line) is close. The value of MAP optimized point is obtained from PMF model. That the value of parameters of Bayesian estimation is close to the MAP estimation demonstrates that the expectation rating is valid and the estimation value of parameters is convergence. 

\subsection{Analysis of Hyperparameters (RQ5)} 
\begin{figure}[htbp]
  \centering
  \includegraphics[width=\linewidth]{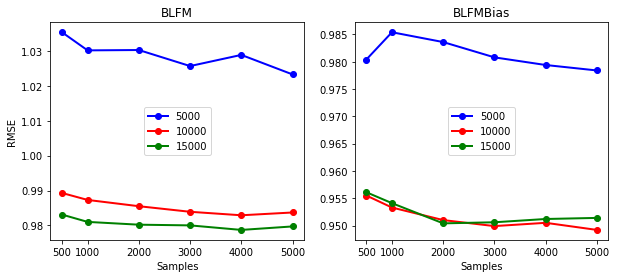}
\caption{Setting the number of fit iterations and samples.}
  \label{fig:5}
\end{figure}

The objective of training BLFM is to fit the best approximated density to the target posterior density. Based on a moderate probabilistic model, we observe the training hyperparameters: the number of iterations and the number of samples. By comparing the blue line (iterations of 5000) with red line (iterations of 10000), green line (iterations of 15000) in Figure \ref{fig:5}, we can demonstrate that the number of iterations increased can significantly improve the performance, but the effect gradually weak. Figure \ref{fig:5} also shows that the number of samples weakly affect the performance (x axis denotes the number of samples). This experiment shows that our model can achieve good performance without continuous enlarging the number of iteration. In fact, our model has comparable run time to an efficiently implemented SVD. 

\section{Discussion and Related Work}
Based on conducting extensive experiments, let us first summary the characteristic of our model, as shown in Table \ref{Tb:4}. 
\begin{table}[htbp]
\renewcommand\arraystretch{1.5}
\caption{The characteristic of the state-of-the-art models}
\begin{center}
\begin{tabular}{ |c|c|c|c|c| } 
\hline 
\textbf{Index} & \textbf{SVD} & \textbf{PMF} & \textbf{BPMF} & \textbf{\textit{BLFM}} \\
\hline 
PM & MAP & MAP & Bayesian & \textit{Bayesian} \\
RG & Parameter & Parameter & Prior & \textit{Prior} \\
RP & medium & lower & higher & \textit{medium} \\
FS & Good & Bad & Bad & \textit{Good} \\
\hline 
\end{tabular}
\end{center}\label{Tb:4}
\end{table}
\begin{itemize}
    \item \textbf{PM} denotes the method of parameter estimation. There have two types of PM, MAP (Optimal Estimation) and Bayesian (Expectation Estimation). Compared to the SVD and PMF with MAP method, the BPMF and BLFM applied Bayesian method achieves better performance. The experiment of RQ1 can demonstrate this. The experiment of RQ4 also demonstrate the validity of expectation estimation. 
    \item \textbf{RG} denotes the method of regularization, includes introducing prior constraint and placing parameter control. On alleviating overfitting, the experiment of RQ3 demonstrate the effectiveness of Bayesian method which includes expectation estimation and prior constraint.
    \item \textbf{RT} denotes the run time for achieving result of Table \ref{Tb:2}. The experiment of RQ1 demonstrates that BLFM with VI is faster than BPMF applied MCMC. The experiment of RQ5 demonstrates that the run time of training BLFM model is mainly decided by the number of iterations.
    \item \textbf{FS} denotes flexibility and scalability of model. Due to the model trained on observed rating entry, the experiment of RQ2 demonstrate that BLFM can incorporate biases into model to enhance the performance. 
\end{itemize}

With rating prediction and LFM method on CF have been widely researched in early literature \cite{gogna2017balancing,lin2019personalization,wang2008unified,rennie2005fast}, a variety of probabilistic factor models \cite{strahl2019scalable,zhang2019probabilistic,monti2018unified,mnih2008probabilistic} and Bayesian method \cite{evans2019scalable,chen2019data,kampman2019variational,lim2007variational,salakhutdinov2008bayesian,harvey2011bayesian} have been proposed to enhance the performance. For estimating parameters of probabilistic model, Expectation Maximization (EM) \cite{ahn2019iterative,gillenwater2014expectation,srebro2003weighted} and MAP were used in early works. Zhang \textit{et al} introduce a novel adaptation of the EM algorithms to learn the parameters of a prediction model for personalised content-based prediction\cite{zhang2007efficient}. Stern \textit{et al} instead used Expectation Propagation and Variational Message Passing to learn a model using both ratings data and content \cite{stern2009matchbox}. Related to applying Bayesian method to solve the probabilistic factor model, the early pioneer works are described as follows:
\begin{itemize}
\item Salakhutdinov \textit{et al}. \cite{salakhutdinov2008bayesian} proposed BPMF based on MCMC and rating matrix which we have discussed detail in experiments.
\item  Lim \textit{et al}. proposed a Bayesian approach to alleviate overfitting in SVD, where priors are introduced and all parameters are integrated out using VI \cite{lim2007variational}. For parameter estimation of low-rank matrix decomposition, this work demonstrated that Bayesian approach is more robust against overfitting than EM and MAP. This work also showed that the effectiveness of VI is not just due to the priors introduced in probabilistic model, but also due to expectation rating. It is trained like as BPMF on rating matrix with extreme data sparsity, this is difference with our work. 
\item  Harvey \textit{et al}. presented a Bayesian latent variable model for rating prediction that models ratings over each user's latent interests and also each item's latent topics\cite{harvey2011bayesian}. This work constructed a probabilistic topic model with assuming the user-interest and item-topic variables drawn from multinomial distributions. It is different with probabilistic factor model in which map users and items to a joint latent space. This work used Gibbs sampling \cite{griffiths2004finding} which is a MCMC method to estimate its parameters and showed that it is competitive with the gradient descent SVD methods. In fact, the model of this work used the ability of Latent Dirichlet Allocation (LDA) \cite{blei2003latent} to extrude latent factors from sparse data. By incorporating user-dependent and item-dependent biases into the model, this work make an extension to enhance the performance of rating prediction. We concede that the ideas of performing rating prediction using Bayesian method and the inclusion of user and item biases is motivated by this work. However, the probabilistic model constructed and Bayesian inference method applied is different.
\end{itemize}

In summary, given observed rating, our BLFM conduct expectation rating with Bayesian method by introducing prior on latent factors of users and items, and have the characteristic of implementing simply, running flexibly, and easy of scalability. Furthermore, we illustrate the characteristics of our model, comparing with the state-of-the-art models. In the future work, we present a exploring research with incorporating additional information \cite{anelli2019local,koren2009collaborative} into BLFM to solve cold start problem \cite{uyangoda2018user,guo2015trustsvd}. By combining prior constraint and posterior expectation estimation, BLFM is more robust against overfitting. In addition, how to choose the prior distribution for factor vectors and biases should be taken application and dataset into consideration. 

\section{Conclusion and Future Work}
In this paper, given the observed rating, we propose probabilistic factor model with the method of Bayesian parameter estimation to implement LFM. First, we introduce Gaussian prior distribution for latent factors of each user and item. Second, we build likelihood function to establish the link between latent factors and observed rating. After constructing the probabilistic model, we apply VI which is a method of Bayesian approximation inference to make expectation rating prediction. We also extend our model by incorporating biases of users and items to enhance the performance. By extensive experiments on Movielens dataset, we demonstrate the effectiveness of our methods. 

In our future work, the model can be extended to include more information, such as the time when each rating was made, the attributes of users and items to tackle cold start problem. Combining the additional information with explicit feedback and implicit feedback is a hot research of CF. The early works applied point estimate for parameters. By our experiment, we hold that the Bayesian method could be applied to solve the LFM model with additional information. The probabilistic model of matrix factorization with Bayesian treatment can be extended to other pattern recognition problems.






%

\bibliographystyle{IEEEtran}
\bibliography{main}

\end{document}